\documentclass[12pt]{article}

\usepackage{scicite}
\newcommand{\CVS}{CsV$_3$Sb$_5$ }
\usepackage{times}
\usepackage{graphicx}
\usepackage{amsmath}
\usepackage{float}

\newenvironment{sciabstract}{%
\begin{quote} \bf}
{\end{quote}}

\title{Direct Observation of Collective Modes of the Charge Density Wave in the Kagome Metal \CVS}

\author
{Doron Azoury,$^{1\dagger}$ Alexander von Hoegen,$^{1\dagger}$ Yifan Su,$^{1}$ Kyoung Hun Oh,$^{1}$\\ Tobias Holder,$^{3}$ Hengxin Tan,$^{3}$ Brenden R. Ortiz,$^{2}$ Andrea Capa Salinas$^{2}$\\ Stephen D. Wilson,$^{2}$ Binghai Yan,$^{3}$ Nuh Gedik,$^{1\ast}$\\
\\
\normalsize{$^{1}$Department of Physics, Massachusetts Institute of Technology, Cambridge, Massachusetts, USA}\\
\normalsize{$^{2}$Materials Department, University of California, Santa Barbara, California 93106, USA}\\
\normalsize{$^{3}$Department of Condensed Matter Physics, Weizmann Institute of Science, Rehovot, Israel}\\
\\
\normalsize{$^\dagger$These authors contributed equally to this work.}\\
\normalsize{$^\ast$To whom correspondence should be addressed; E-mail:  gedik@mit.edu.}
}

\date{}

\begin{document} 


\baselineskip24pt


\maketitle


\begin{sciabstract}
  A new group of kagome metals AV$_3$Sb$_5$ (A = K, Rb, Cs) exhibit a variety of intertwined unconventional electronic phases, which emerge from a puzzling charge density wave phase. Understanding of this parent charge order phase is crucial for deciphering the entire phase diagram. However, the mechanism of the charge density wave is still controversial, and its primary source of fluctuations – the collective modes – have not been experimentally observed. Here, we use ultrashort laser pulses to melt the charge order in \CVS and record the resulting dynamics using femtosecond angle-resolved photoemission. We resolve the melting time of the charge order and directly observe its amplitude mode, imposing a fundamental limit for the fastest possible lattice rearrangement time. These observations together with ab-initio calculations provide clear evidence for a structural rather than electronic mechanism of the charge density wave. Our findings pave the way for better understanding of the unconventional phases hosted on the kagome lattice.
\end{sciabstract}

\begin{figure}
\begin{center}
  \centering{\includegraphics*[width=0.85\columnwidth]{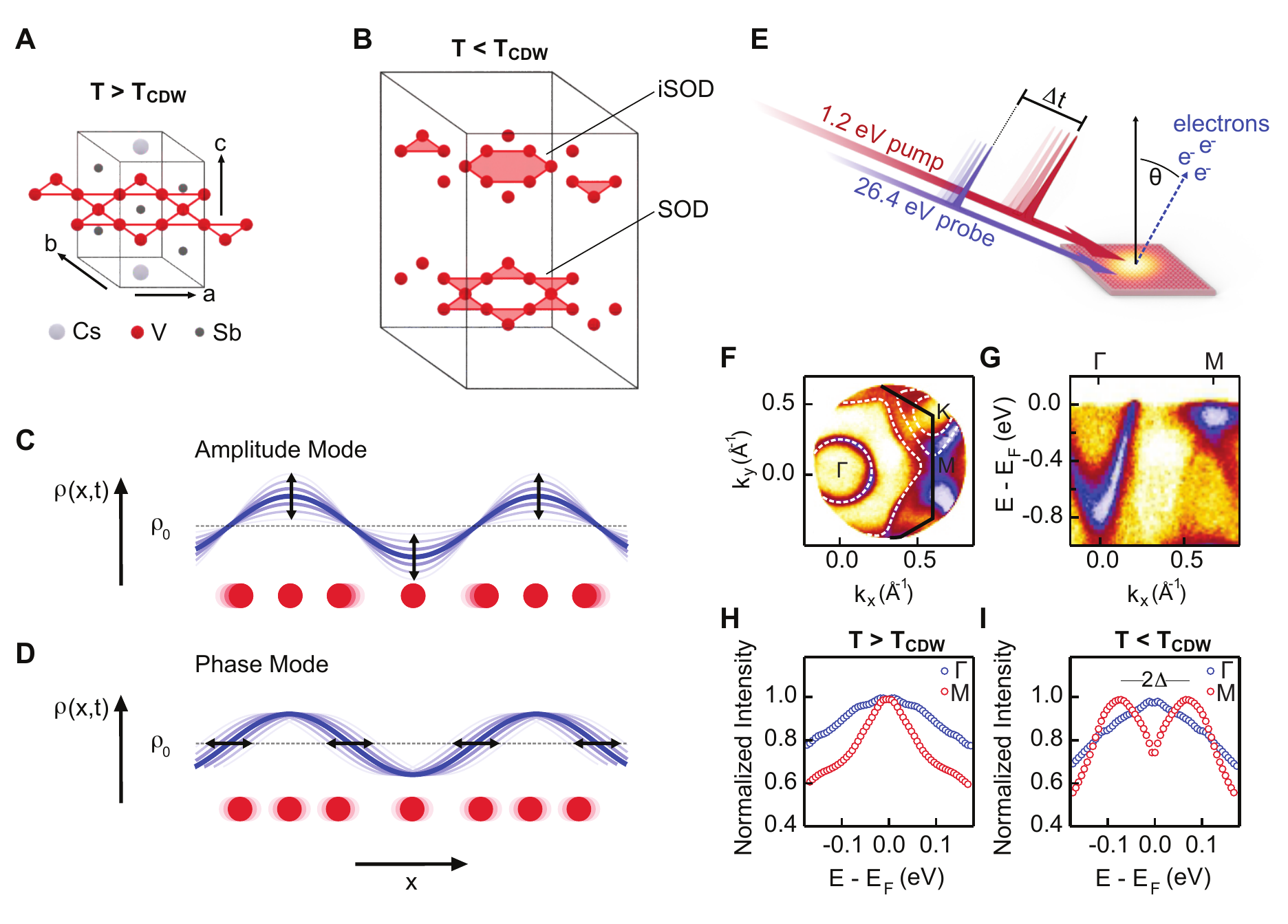}}
  \caption{{\bf Charge density wave in \CVS.} ({\bf A}) Crystal structure of \CVS above the CDW transition temperature $T_{\text{CDW}}$. ({\bf B}) Schematic of the two stacked distortion types in \CVS, Star-of-David (SOD) and inverse Star-of-David (iSOD) in the charge denisty wave phase belwo $T_{\text{CDW}}$. ({\bf C} and {\bf D}) Illustration of the collective motion induced by amplitude and phase modes in a CDW system, where the red spheres and the blue curves correspond to the lattice and charge density, respectively. ({\bf E}) Schematic description of the tr-ARPES experiment. A synchronized pump-probe pulse pair excite the sample, and the emitted photo-electrons are detected by a time-of-flight electron detector that resolves their energy and momentum. ({\bf F} and {\bf G}) Static ARPES spectra, showing a constant energy cut at the Fermi level (F) and energy-momentum cut along the $\Gamma$-M high symmetry line (G). The characteristic spectral features of \CVS are identified at the $\Gamma$ point (parabolic band), M point (Van Hove singularity) and the Fermi surface shows good agreement with a superimposed DFT calculation in (white dashed line in (F)). ({\bf H} and {\bf I}) Symmetrized energy density curves, extracted at the $\Gamma$ and M points, above and below $T_{\text{CDW}}$, showing the opening of a gap $2\Delta$ at the M point.}
  \label{fig1}
\end{center}
\end{figure}

Highly correlated systems may exhibit a complex electronic phase diagram, where multiple phases are intimately related, and can be thought of as intertwined phases. In some cases, such as high-temperature superconductors for example~\cite{fradkin2015colloquium}, understanding the nature of individual phases is insufficient, and knowledge of the transition from one phase to another is crucial in order to understand the physics of these systems. Such an array of intertwined phases were recently discovered in a new class of vanadium-based kagome metals AV$_3$Sb$_5$ (A = K, Rb, Cs)~\cite{ortiz2019new,ortiz2021fermi,jiang2021unconventional,liang2021three,ortiz2020cs,zhao2021cascade,chen2021roton,ortiz2021superconductivity,wang2021charge}. In these compounds all the phases emerge from a parent phase of a charge density wave (CDW). In the most simple case, a CDW forms when regions of the Fermi surface with high density of states can be connected by the same nesting wavevector. This leads to a spatial modulation of the electron density and concomitant distortion of the crystal lattice (see Fig.~\ref{fig1}C), which lowers the crystal’s energy and opens an energy gap. Recent studies have revealed the properties of the CDW phase in AV$_3$Sb$_5$~\cite{ortiz2021fermi,jiang2021unconventional,liang2021three,ortiz2020cs,zhao2021cascade,chen2021roton}. Specifically, the unit cell doubles along both in-plane directions, with nesting vectors connecting the high density of states at the van Hove singularities, located at the M points on the Brillouin zone boundary. This is also where the CDW gap opens, while the band structure near the $\Gamma$ point remains gapless~\cite{hu2022rich}. The CDW in $AV_3Sb_5$ was shown to be very amenable to external perturbation such as hydrostatic pressure~\cite{wang2021charge,du2021pressure,chen2021double,yu2021unusual} or doping~\cite{oey2022fermi} and recent studies revealed an additional weak reconstruction along the out-off-plane direction~\cite{liang2021three,stahl2022temperature,xiao2022coexistence}. Additionally, the observations of chirality in the CDW phase~\cite{shumiya2021intrinsic,jiang2021unconventional,wang2021electronic}, the absence of acoustic phonon softening~\cite{li2021observation} and a pronounced anomalous Hall effect~\cite{yu2021concurrence,liu2022observation,yang2020giant} suggest an electronic origin of the charge order in $AV_3Sb_5$. However, the underline mechanism that forms the CDW was not directly resolved so far. In order to better understand the complete broken symmetry phase diagram, and how individual phases relate to one-another, determining the parent phase mechanism is a crucial step.

To address these open questions, time-resolved studies were proved to be a powerful tool to resolve otherwise inaccessible aspects of interactions by measuring a response to impulsive excitation. These techniques are sensitive to collective dynamics of a condensed matter system, including the melting of long-range order and the fundamental modes of a CDW~\cite{zhu2018unconventional,zong2019evidence,zong2019dynamical}. The latter manifest as collective oscillations of the electronic density modulation, or a sliding motion of the CDW, known as the amplitude and phase modes, respectively (see illustration in Fig.~\ref{fig1}C and D). These modes typically constitute the lowest energy collective excitations (GHz to few THz) of the system. As such, they define a distinctive time scale, imposing a minimum value of the structural response time in a photo-induced phase transition typically on the order of a few hundred femtoseconds. Conversely, if the melting is purely electronic it takes place on a much faster, electronic, femtosecond time scale. Therefore the combined knowledge of the response time and the collective modes allow the identification of the dominant interaction driving the phase transition. Experimentally, amplitude and phase modes in CDWs are routinely observed using Raman and infrared spectroscopy, yet, eluded detection in $AV_3Sb_5$ so far~\cite{liu2022observation,wu2022charge,ratcliff2021coherent,wang2021unconventional,wu2021large}. To accurately describe and predict the phase-transitions in any condensed matter system a comprehensive knowledge of the collective modes is required, leaving a large gap in our understanding of the $AV_3Sb_5$ kagome system. 

Here, we perform time- and angle-resolved photoemission spectroscopy (trARPES) to study the dynamics of the CDW in \CVS. trARPES gives access to the non-equilibrium electronic band structure during a photo-induced phase transition on a femtosecond (fs) timescale. This time, energy and momentum resolved approach allows us to directly probe the CDW gap, and observe non-thermal melting of the charge order as well as its ultrafast recovery~\cite{perfetti2006time,schmitt2008transient,rohwer2011collapse}. The ultra-short photoexcitation impulsively excites coherent oscillations throughout the electronic band-structure, which agree well with known CDW induced phonons. Yet, trARPES unfolds the momentum integrated information obtained in other techniques, allowing for the assignment of band-selective electron-phonon coupling~\cite{yang2019mode,gerber2017femtosecond,sakamoto2022connection}. Analysis of the time dependent signal within the CDW gap at the M-point reveals two additional modes, so far unobserved by any other spectroscopic method, which differ significantly from the detected phonons by showing a frequency softening at elevated fluences. In combination with ab initio density functional theory (DFT) calculations we identify these modes as the amplitude modes of the three-dimensional CDW in \CVS. Finally, the measurement of the charge-order melting time together with the detection of the amplitude modes establish a complete basis to determine the interaction that dominates the CDW phase in \CVS as structural.

\CVS exhibits a quasi-two-dimensional crystal structure (space group P6/mmm) of alternately stacked alkali and V–Sb planes. Each V–Sb plane hosts a two-dimensional kagome network of V atoms in which the hexagonal centers are occupied by Sb atoms. The additional out-of-plane Sb atoms form two honeycomb lattices above and below the V kagome network (see Fig.~\ref{fig1}A). Below a temperature of $T_{CDW}=94K$, the crystal undergoes a CDW phase transition and the unit cell doubles along all in-plane directions. Recent diffraction and static ARPES studies suggest an additional lattice reconstruction along the out-of-plane direction: the lattice relaxes into a mixture of 2x2x2 and 2x2x4 reconstructions through a so-called star-of-David (SOD) and inverse star-of-David (iSOD) distortions of the Vanadium atoms on adjacent layers (see Fig.~\ref{fig1}B)~\cite{liang2021three,xiao2022coexistence,ortiz2021fermi}. Static ARPES measurements revealed the equilibrium band-structure showing features associated with the kagome network: Dirac fermions at the K point and Van Hove singularities at the M point of the Brillouin zone. The CDW phase transition renormalizes the band-structure and opens a highly anisotropic gap around the M point. Fig.~\ref{fig1}F shows a static ARPES measurement of the Fermi surface in \CVS, taken at $T=35K<T_{CDW}$, as captured by our time-of-flight based ARPES detection (see Supplementary Material). The high symmetry points ($\Gamma$, K and M) are marked on two different cuts (see Fig.~\ref{fig1}F and G). Energy distribution curves (EDC), extracted at the $\Gamma$ and M points (see Fig.~\ref{fig1}H and I) show the opening of a gap at the M point while the $\Gamma$ band remains metallic. Our static ARPES data is consistent with all previous studies~\cite{wang2021distinctive,nakayama2021multiple,luo2021distinct,lou2022charge}. 

\begin{figure}
\begin{center}
  \centering{\includegraphics*[width=0.8\columnwidth]{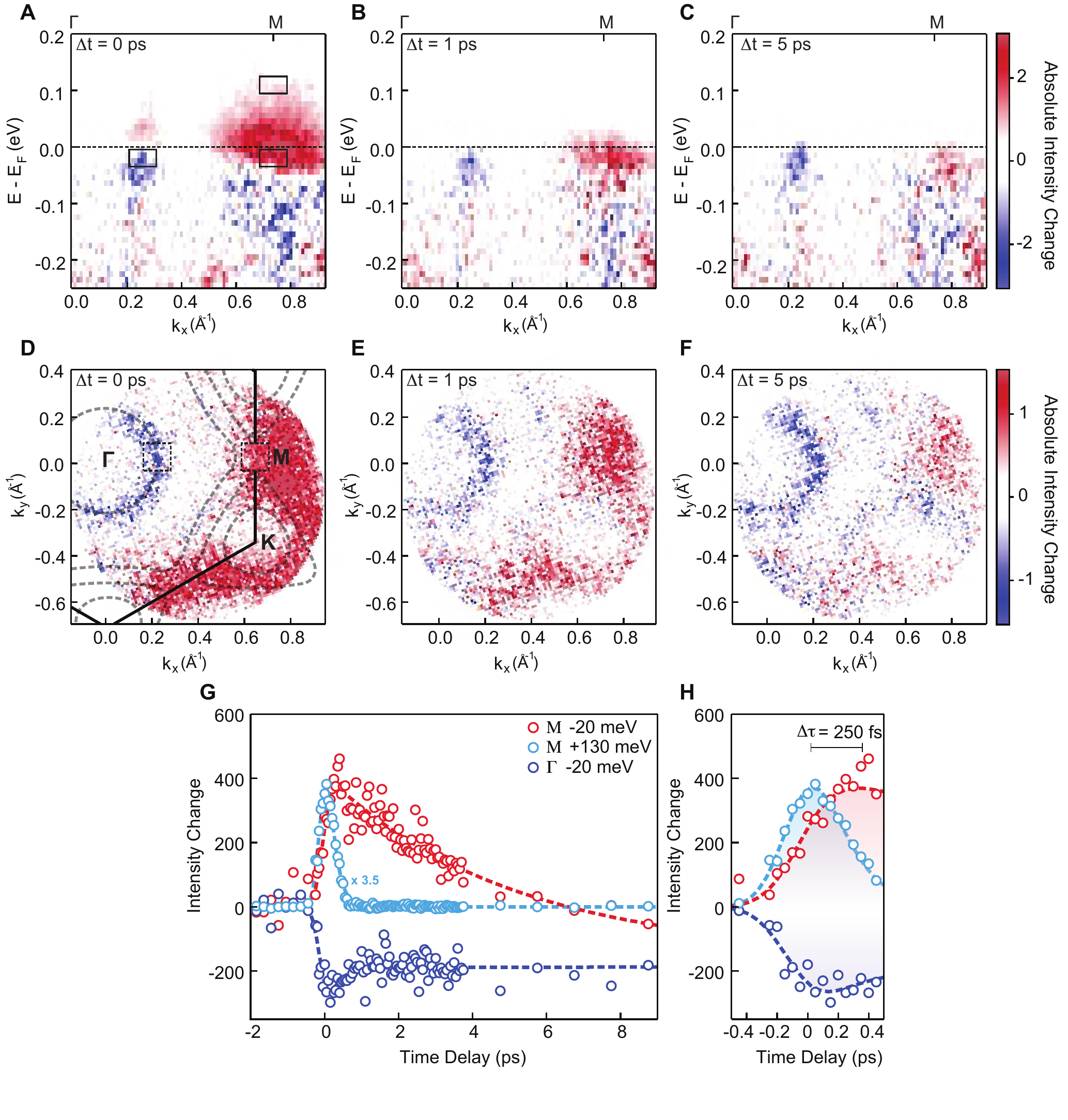}}
  \caption{{\bf Charge density wave dynamics in \CVS.} ({\bf A-F}) Differential spectra for an energy-momentum cut along the $\Gamma$-M high symmetry line (A-C) and a constant energy cut at $E=E_{\text{F}}-20meV$ (D-F), for three different delays (t = 0ps, 1ps, 5ps). The spectral gain (loss) signal is marked in red  (blue). The transition between loss and gain signal at the $\Gamma$ metallic band is symmetric with respect to the Fermi level (dashed black line in (A-C)), whereas the  gap region at the M point shows significant gain below the Fermi level. In (D-F) the gain signal allows the identification of the bands that host the CDW gap. A sketch of the band-structure is superimposed for  reference. ({\bf G}) Spectral intensity as a function of time delay, extracted from the regions marked in (A). The metallic $\Gamma$ band (dark blue) shows a fast depletion followed by partial recovery. The excited states at the M point (turquoise) show a fast rise, synchronized with the depletion peak of the $\Gamma$ band, and a fast decay. The in-gap intensity at the M point (red) show a slower rise time, corresponding to the melting time of the CDW, followed by a 4 ps recovery time. The dashed lines correspond to a double-exponential fit. ({\bf H}) A close-up over the onset of the interaction, showing a 250 fs delay of the in-gap signal.}
  \label{fig2}
\end{center}
\end{figure}

To investigate the CDW in \CVS, beyond what is possible within the static approach, we use an ultrashort near-infrared (1.2eV) pump pulse to transiently perturb the CDW order and a second, time delayed, extreme ultra-violet (26.4 eV) probe pulse to stimulate photoemission of electrons (see Fig.~\ref{fig1}E). We scan the delay ($\Delta t$) between the pump-probe pair, and perform an ARPES measurement for each delay, to obtain the non-equilibrium electron band structure during the photo-induced phase transition. To effectively visualize the transient changes of the band-structure, we treat the unperturbed spectra at negative delay as a baseline and plot the relative changes in blue (negative) and red (positive) for selected time delays ($\Delta t$ = 0 ps, 1 ps and 5 ps) in Fig.~\ref{fig2}.  Fig.~\ref{fig2}A-C shows the relative changes along the energy-momentum cut between the $\Gamma$ and M points including both the metallic $\Gamma$-band and the CDW gap at M. As expected from a metallic band, the transition between the loss and gain signal at the $\Gamma$ point is found at the Fermi energy ($E_{\text{F}}$) as a result of the transiently increased electronic temperature following the laser excitation. At the M point, the transition between loss and gain is shifted to below the Fermi energy. This is the transient closing of the CDW gap and filling in of the previously unoccupied states inside the gap region, leading to spectral gain below the Fermi energy~\cite{maklar2021nonequilibrium}. Following the intensity contour therefore allows us to visualize the energy-momentum cross section of the CDW gap itself. Above $E_{\text{F}}$, at both $\Gamma$ and M points we observe the population of excited states, which completely decay within 1ps (see Fig.~\ref{fig2}A and B), while the gap remains closed, and reopens at later times (see Fig.~\ref{fig2}B and C). The constant energy cuts (E = -20 meV), spanned by the two orthogonal momentum directions $k_x$ and $k_y$ shown in Fig.~\ref{fig2}D-F, provide complementary information. Here, we can directly identify the bands that host the CDW gap by using differential contrasting: a gain signal is associated with the gap related bands, whereas a loss signal is associated with a normal metallic band. We repeated the trARPES measurements above the equilibrium transition temperature at $T = 110 K > T_{CDW}$ and observed a uniform metallic response throughout the band structure (see Supplementary Material). A recent classification of the bands that host the CDW gap in \CVS shows consistent results with Fig.~\ref{fig2}D-F, distinguishing between three bands of which two host a substantial gap~\cite{kang2022twofold}. 
 
Next, we extract the spectral intensity in specific regions of interest (see rectangles in Fig.~\ref{fig2}D) at the top of the $\Gamma$ and M bands (E = - 20 meV) and in the excited states region (E = 130meV) at the M point as a function of time delay (see Fig.~\ref{fig2}G). Due to their prompt excitation and short lifetime, we identify the temporal overlap ($t_0$) of pump and probe with the peak of the excited states signal. Clearly, the response at the M and $\Gamma$ points below $E_F$ is very different – the metallic band at the $\Gamma$ point is depleted nearly instantaneously, whereas inside the gap at the M point we detect an intensity increase. The long-lived depletion at the $\Gamma$ point can most likely be related to the limited recombination channels for this particular band (see Supplementary Material), similarly seen as an anomalously long recovery behavior in time-resolved reflectivity measurements~\cite{ratcliff2021coherent}. The in-gap intensity decays in about 4ps, indicating that the CDW gap reopens on a similar timescale as observed in other compounds~\cite{eichberger2010snapshots}. Fig.~\ref{fig2}H shows a zoom-in around $\Delta t=0$ to better visualize the behavior during the photo-excitation. The excited states at the M point (turquoise) show a fast rise, synchronized with the depletion peak of the $\Gamma$ band. By contrast, the in-gap intensity clearly shows a 250fs delayed response compared to the peak in the excited states, providing a direct measurement of the non-thermal CDW melting time. Previous studies of other systems used a discriminatory approach~\cite{hellmann2012time}, based on the CDW melting time, to distinguish between CDWs dominated by either electron-electron or electron-phonon interactions. A key ingredient to this approach is the comparison between the melting time and the fundamental time-scale imposed by the dominant interaction, which can range between a few (electronic) to hundreds (structural) of femtoseconds. For example, the electronically driven CDW in $1T-TiSe_2$ was observed to melt within about 40fs~\cite{rohwer2011collapse}, while the structural CDW in $1T-TaS_2$ melts within 230fs~\cite{hellmann2012time}. The measured melting time of 250fs in \CVS points towards electron-phonon interactions dominating the CDW in \CVS, as it corresponds to the typical time scale of atomic motion in a solid. Nevertheless, a complete determination of the CDW mechanism is only possible with precise knowledge of a reference time-frame, i.e., the time it takes for the atoms to move back to their undistorted arrangement. This time-scale is given by the amplitude mode frequency $\Omega_\Delta$, which so-far eluded detection in \CVS\cite{liu2022observation,wu2022charge,ratcliff2021coherent,wang2021unconventional,wu2021large}.

\begin{figure}
\begin{center}
  \centering{\includegraphics*[width=0.6\columnwidth]{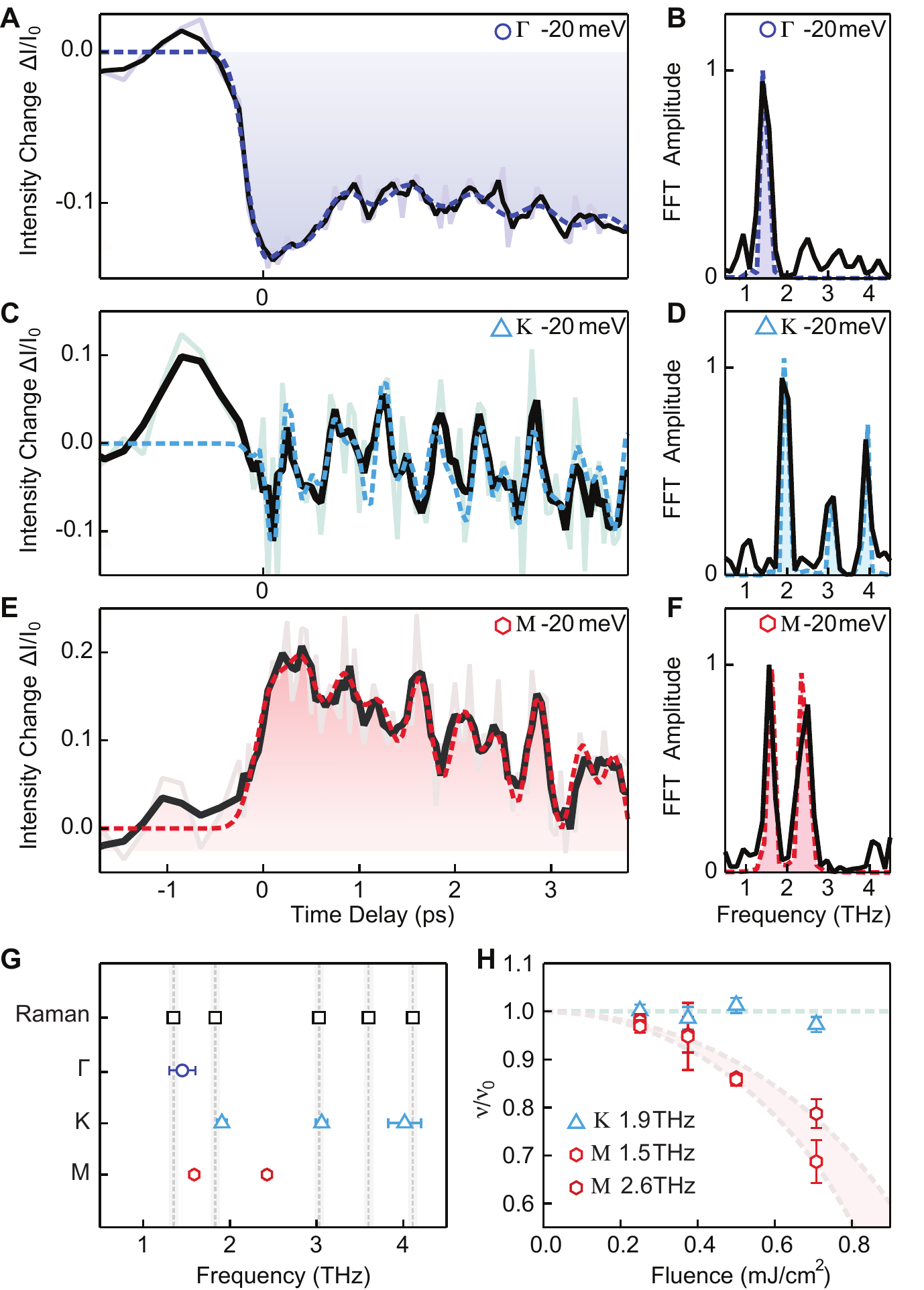}}
  \caption{{\bf Coherent phonons and charge density wave collective modes in \CVS.} ({\bf A-D}) Time-resolved intensity change at the $\Gamma$ (A) and K (C) bands for an excitation fluence of 0.5 $mJ/cm^2$. The faded line is the raw data, the black line is the data smoothed by a three-point average and the dashed line is a fit to the data (see Supplementary Material). (B) and (D) show Fourier intensity of the raw signal (black line) after subtracting the slowly varying background and the Fourier Intensity of the oscillatory part of the fit (dashed line) in (A) and (C), respectively. ({\bf E} and {\bf F}) In-gap Intensity change as a function of time delay also for an excitation fluence of 0.5 mJ/cm2. The same analysis as in (A) and (C) reveal two dominant frequencies, at 1.6 THz and 2.4 THz. ({\bf G}) Comparison between the frequencies of the coherent oscillations at the high symmetry points and known Raman phonons~\cite{wu2022charge}. The coherent modes at the K and $\Gamma$ bands agree well with the Raman modes that appear below $T_{\text{CDW}}$. The coherent modes at M do not coincide with any known Raman mode. ({\bf H}) Relative frequency shift of the 1.9 THz phonon mode and the in-gap modes as a function of pump fluence. The phonon mode frequency remains constant while the in-gap modes show softening. The dashed lines are a guide to the eye, and the error bars correspond to the uncertainties of the time-domain fit.}
  \label{fig3}
\end{center}
\end{figure}

This missing information can be retrieved by examining the coherent oscillations which are superimposed on our time-resolved spectral intensity signal. We make use of our momentum resolution to examine these band-selective oscillations at specific high symmetry points of the Brillouin zone. Fig.~\ref{fig3}A, C and D show the transient spectral intensity change of the bands close to $\Gamma$, K and M just below $E_{\text{F}}$. A Fourier analysis of the coherent oscillations reveals well separated peaks significantly above the noise floor of the experiment. Most strikingly, the distinct bands at $\Gamma$ K and M show different oscillation frequencies (see Fig.~\ref{fig3}B, D and F). We note that the intensity increase at negative delays are due to a pump-polarization dependent space charge effect, which does not affect the positive delay signal (see Supplementary Material for details). A comparison with previous Raman~\cite{wu2022charge} and coherent phonon spectroscopy~\cite{ratcliff2021coherent} measurements, shows good agreement between the detected oscillation frequencies at $\Gamma$ and K and the reported phonon modes of \CVS (see Fig.~\ref{fig3}G). Importantly, all common phonon modes that appear in both Raman and our trARPES measurements are associated with the CDW phase transition. We further find good agreement with frozen-phonon DFT calculations which predict pronounced electron-phonon coupling to the high symmetry points of the electronic band-structure (see Supplementary Material). Interestingly, an equivalent analysis of the in-gap intensity at the M point reveals two collective modes at 1.6 THz and 2.4 THz (see Fig.~\ref{fig3}F), which do not appear in any static or time-resolved measurement. These modes do not overlap with any of the phonon modes observed at other momentum points (see Fig.~\ref{fig3}B and D). Finally, a fluence dependent measurement reveals a pronounced softening of these two modes, in stark contrast to the phonons, which hardly respond to the pump fluence variation (see Fig.~\ref{fig3}H). These observations prompt the question of the origin of these in-gap modes.

We recall that the amplitude mode renormalizes the size of the CDW gap and can consequently be observed as coherent oscillations of the in-gap intensity~\cite{gruner1988dynamics}. Thus, the measured oscillations at the M point can, in principle, be assigned to coherent oscillations of the amplitude mode. Yet, the observation of two modes modulating the in-gap intensity cannot be reconciled with amplitude oscillations of a single complex order parameter. Recent studies have shown that the out-of-plane lattice reconstruction consists of alternating SOD and iSOD distortions on adjacent layers~\cite{kang2022charge} (see Fig.~\ref{fig1}B). Therefore, in the limit of vanishing coupling between the layers, one can regard the three-dimensional CDW as a superlattice of stacked SOD and iSOD CDWs each with their individual order parameter and set of collective modes. Indeed, recent x-ray diffraction measurements reveal that the 3-D structure is very amenable to perturbation suggesting only weak coupling between the layers~\cite{kautzsch2022incommensurate}. As outlined above, our photo-excitation quenches the CDW and accordingly suppresses the weak coupling among the SOD and iSOD distorted layers allowing them to perform amplitude oscillations at their own eigenfrequencies. To estimate these frequencies, we calculated the free energy for the two distortions using DFT methods, which allowed us to extract the two associated amplitude mode frequencies (see Fig.~4 and Supplementary Material) $\nu_{SOD}=3.2$ THz  and $\nu_{iSOD}=4.6$ THz for the SOD and iSOD distortions, respectively. These calculations represent a maximum value as they correspond to T=0, however, the ratio $\alpha=\nu_{iSOD}/\nu_{SOD}$ of the two frequencies is unaffected by this uncertainty and we use it as a quantitative metric to compare with our experimental result. Considering the complex lattice and electronic structure of \CVS, we find surprisingly good agreement between the experimental $\alpha=1.5\pm0.1$ and theoretical $\alpha=1.4$ frequency ratio. Therefore, we find that the coherent oscillations within the gap can be ascribed to independent amplitude oscillations of the SOD and iSOD layers. In principle, each distortion type generates a specific energy gap, however since the distortion magnitude is very similar in both cases (See Fig. 4), their associated gaps are almost degenerate (see Supplementary Material), and therefore cannot be resolved independently within our measurement energy resolution. Our findings can offer a possible reason for the absence of the amplitude modes in Raman measurements. The structural distortions associated with the amplitude modes of the alternating SOD and iSOD layers are exactly the opposite of each other and may therefore render this mode silent for a typical Raman experiment. Our trARPES experiment on the other hand is directly sensitive to changes of the CDW gap and therefore allows us to identify the amplitude modes according to their coherent beating pattern. These modes provide the fastest possible time scale for the structural reconfiguration of a CDW – melting cannot occur faster than half an oscillation cycle $T_\Delta=\frac{1}{\nu_\Delta}$ of the amplitude mode. Accordingly, we compare the amplitude mode frequencies of the two distortions and we find $T_\Delta\approx$ 200 fs and 300 fs for the SOD and iSOD distortion, respectively. This result is in excellent agreement with the observed melting time $\tau\approx$250 fs and clearly set electron-phonon interactions as the dominant mechanism that drives the CDW in \CVS.

Apart from clarifying important aspects of the CDW order itself, the collective excitations in the low THz-regime presented here may be furthermore valuable in understanding the superconducting phase in \CVS which develops below $T=$3 K. The superconducting order parameter is believed to form a commensurate pair density wave, possibly of  unconventional origin~\cite{chen2021roton}, and the phase diagram both as a function of doping~\cite{kautzsch2022incommensurate} and pressure~\cite{chen2021double,zheng2022emergent} strongly suggests that the emergence of superconductivity is linked to the parent CDW phase, and might even compete with it. As \CVS does not feature any magnetic order, the collective excitations associated with the CDW are therefore a prime candidate for the possible pairing glue if the superconducting instability is unconventional. Based on our findings, we believe that a detailed analysis of this possibility is called for.

To conclude, we have used trARPES to study the CDW in \CVS and observed its non-thermal melting and recovery. From this time-resolved measurement, we were able to extract the CDW melting time and identify coherent oscillations of the CDW amplitude mode, which provide a complete basis to identify the dominant interaction in the CDW phase. We find that the observed melting time of $\tau\approx250fs$, together with an amplitude mode frequencies are evidence that the CDW phase is dominated by electron-phonon interactions. This strong evidence of a structural nature of the CDW in \CVS unambiguously demonstrates the important role the crystal lattice plays in a system that is believed to be dominated by electronic correlations \cite{shumiya2021intrinsic,jiang2021unconventional,wang2021electronic,li2021observation,yu2021concurrence,liu2022observation,yang2020giant}.

\bibliography{CollectiveModeCVS}

\bibliographystyle{Science}

{\bf Acknowledgments}: The authors thank Dongsung Choi, Baiqing Lyu and Masataka Mogi for technical support. The work at MIT was supported by the US Department of Energy, BES DMSE (data taking, analysis and manuscript writing) and Gordon and Betty Moore Foundation’s EPiQS Initiative grant GBMF9459 (instrumentation). D. A. acknowledges financial support by the Zuckerman STEM Leadership Program. A.v.H. gratefully acknowledges funding by the Humboldt foundation. B.Y. acknowledges the financial support by the European Research Council (ERC Consolidator Grant ``NonlinearTopo'', No. 815869) and the ISF - Personal Research Grant (No. 2932/21). S.D.W., B.R.O., and A.C.S. gratefully acknowledge support via the UC Santa Barbara NSF Quantum Foundry funded via the Q-AMASE-i program under award DMR-1906325. {\bf Author contributions}: N.G. supervised the study. D.A., A.V.H., Y.S. and K.H.O. performed the trARPES measurements. A.V.H. and D.A. analyzed the data. T.H., H.T. and B.Y. performed the DFT calculations. B.R.O., A.C.S. and S.D.W. performed the crystal growth. All authors discussed the results and contributed to the final manuscript. {\bf Competing interests}: Authors declare that they have no competing interests. {\bf Data and materials availability}: All data are available in the main text or the supplementary materials.

\clearpage

\begin{figure}
\begin{center}
  \centering{\includegraphics*[width=0.55\columnwidth]{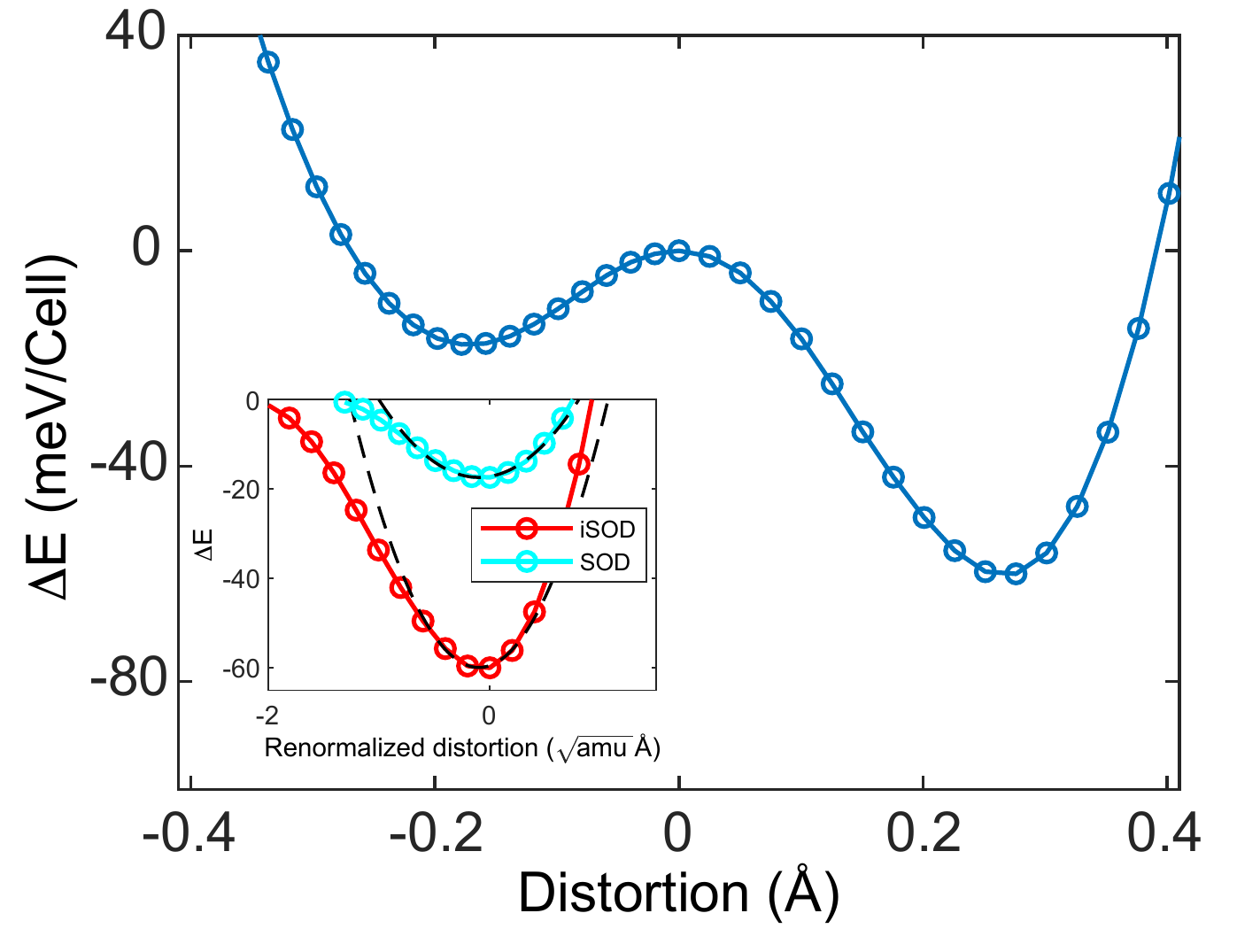}}
  \caption{{\bf Total energy profiles for SOD and iSOD distortions.} DFT calculation of the total energy difference for the SOD (negative side) and iSOD (positive side) distortions in \CVS with respect to the pristine structure. The distortion parameter is defined by Eq.~1 in the Supplementary Material. (inset) The mass-weighted distortion - energy relation to extract the collective oscillation frequency by Eq.~2 in the Supplementary Material. Parabolic fits are shown in dashed black.}
  \label{fig4}
\end{center}
\end{figure}

\end{document}